# ENERGY-AWARE DISK STORAGE MANAGEMENT: ONLINE APPROACH WITH APPLICATION IN DBMS

Peyman Behzadnia[1], Yi-Cheng Tu[1], Bo Zeng[2] and Wei Yuan[3]

[1]Dept. of Computer Science & Engineering, University of South Florida, Tampa, FL, USA
[2] Department of Industrial Engineering, University of Pittsburgh, Pittsburgh, PA, USA
[3] Dept. of Industrial & Management Systems Eng., Univ. of South Florida, Tampa, FL, USA

## ABSTRACT

*Energy consumption has become a first-class optimization goal in design and implementation of data-intensive computing systems. This is particularly true in the design of database management systems (DBMS), which was found to be the major consumer of energy in the software stack of modern data centers. Among all database components, the storage system is one of the most power-hungry elements. In previous work, dynamic power management (DPM) techniques that make real-time decisions to transition the disks to low-power modes are normally used to save energy in storage systems. In this paper, we tackle the limitations of DPM proposals in previous contributions. We introduced a DPM optimization model integrated with model predictive control (MPC) strategy to minimize power consumption of the disk-based storage system while satisfying given performance requirements. It dynamically determines the state of disks and plans for inter-disk data fragment migration to achieve desirable balance between power consumption and query response time. Via analyzing our optimization model to identify structural properties of optimal solutions, we propose a fast-solution heuristic DPM algorithm that can be integrated in large-scale disk storage systems for efficient state configuration and data migration. We evaluate our proposed ideas by running simulations using extensive set of synthetic workloads based on popular TPC benchmarks. Our results show that our solution significantly outperforms the best existing algorithm in both energy savings and response time.*

## 1. INTRODUCTION

Data centers consume massive and growing amount of energy. A report shows that, in 2013, data centers in the Unites States consumed an estimated 91 billion kilowatt-hours (kWh) of electricity (which costed roughly 7.5 billion US dollars) and are on-track to reach 140 billion kWhs by 2020 [1]. Database management system (DBMS) was found to be the major energy consumer in the software stack of modern data centers [2]. And, among all components of database, storage system is one of the most energy-hungry constituents [2]. Storage system is estimated to consume 25% of total energy consumption in data centers [3, 4, 5] and adding up to 40% of the IT equipment energy consumption [6]. In [2], researchers analyzed a subset of 7 years of online transaction processing (OLTP) systems benchmarked with TPC-C, concluding that the storage system demands more than 70% of power in a TPC-C system. Thus, there is a strong demand for reducing the energy consumption in storage systems. Hence, in this paper, we tackle the problem of designing a power-aware disk storage system in database servers. Note that the use of SSD drives simplifies the problem since they are highly energy efficient compared to HDDs, but, as of today, SSDs are still not in a position to replace all magnetic disks in large-scale storage systems as the primary storage medium, especially those handling today's big data applications, since hard drives have advantages in terms of cost and capacity compared to SSDs [7], [8]. Thus, the focus of this paper is on traditional hard drives and trade-offs between performance and energy efficiency.

Making storage systems green has been addressed in many research efforts in the literature. In previous work, dynamic power management (DPM) algorithms are normally used to save energy in disk storage systems. Such algorithms make real-time decisions on when to transition magnetic disks to lower-power modes with the price of longer response time to data access requests. Many modern hard disks have two power states: active and stand-by. Disks in stand-by mode stop rotation completely thus consume significantly less energy than in active state. However, it incurs a considerable cost in response time and energy to spin up the disk to active mode in order to serve a request. Figure 1 shows the detailed specifications related to the power and transition time among different states of a typical multi-mode disk (model Ultra-star 7k6000 from IBM) [9].

Given the aforementioned penalty cost in response time and energy related to disk state transition, traditional DPM methods provide either little energy savings or suffer from significant performance degradation. More effective DPM techniques attempted to improve this limitation by extending the idle period of disks by either controlling the I/O intervals [4], [10], [11], [12], [13], [14], [15], [16], [17] or consolidating data on subset of disk [18], [19], [20], [21], [22]. The first set of works usually considers single-disk systems and utilizes energy-efficient caching or pre-fetching techniques to prolong the idle periods in the I/O workload. The second set of works basically consolidates the most frequently accessed data (called "hot" data in literature) on subset of disks to allow "cold" disks sleep longer. Therefore, they normally perform corresponding inter-disk data migration in order to achieve the hot data consolidation goal. As the major limitation, work of this type cannot effectively enough handle the dynamic workloads where arrival rate of data requests changes significantly with respect to time. Furthermore, they do not usually provide efficient disk state configuration or inter-disk data migration plans.

In this paper, we tackle the limitations of the previous work. The best known algorithm in the literature that tries to handle dynamic environment is named dynamic block exchange (BLEX) presented in [22]. However, we believe BLEX, again, does not strongly enough adapt to dynamicity in the workload since it performs insufficient inter-disk data migration. We address this issue by introducing an optimization model that integrates model predictive control (MPC) strategy to accommodate dynamic scenarios by enabling optimization actions in an online fashion. In particular, the current control action is obtained dynamically where, at each sampling instant, a finite horizon optimization problem is solved and its optimal solution is applied as the current control decision. Such procedure repeats along the whole control process. Our experimental results clearly show that the proposed model outperforms the BLEX algorithm significantly in terms of both energy savings and response time.

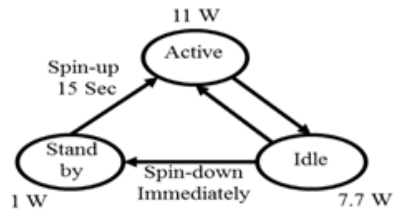

Figure 1. Power modes and their power consumption of the IBM Ultra-Star 7k6000

Furthermore, we propose a heuristic DPM algorithm that can be integrated to large-scale disk storage systems, where the MPC potentially incurs significant overhead, in order to provide fast and efficient power saving solutions. Our experimental evaluations show that the solution achieved by the heuristic algorithm is very close to that of the DPM optimization model in all types of experimental scenarios. Also, the proposed heuristic algorithm outperforms the BLEX algorithm in terms of both response time and energy savings.

## 1.1. CONTRIBUTIONS AND ROADMAP

This paper significantly extends our previous work [23]. In summary, this paper makes the following contributions:

(1) Via analyzing the model to identify structural properties of optimal solution, we propose a fast-solution heuristic DPM algorithm that dynamically determines efficient disk state configuration and inter-disk data migration. It can be integrated to large-scale disk storage systems holding today's big data, where finding the most optimal solution might be long, in order to achieve near optimal power saving solution (near to that of the optimization model) within short periods of computational time;

(2) We significantly extend our experimental simulations using extensive set of synthetic workloads based on popular TPC benchmarks to evaluate our solutions. We compare our experimental results in terms of both power saving and response time with those of the best existing dynamic algorithm named dynamic block exchange algorithm (BLEX). Our results clearly demonstrate remarkable energy savings while satisfying the given performance bound. Our heuristic DPM algorithm and MPC-based optimization model outperforms the BLEX algorithm significantly in terms of both energy savings and response time in data access;

(3) We explore important characteristics of our DPM optimization model and perform experimental evaluations on its important features. Accordingly, analytical results on DPM model characteristics are presented in detail;

The remainder of this paper is organized as follows: Section 2 provides required background and survey on related work in the literature; Section 3 describes the proposed DPM optimization model in detail; Section 4 illustrates our proposed heuristic DPM algorithm; Section 5 presents in detail our experimental methodology; and Section 6 concludes the paper.

## 2. RELATED WORK

Energy conservation in disk-based storage systems is addressed in many research projects. They can be primarily distinguished into two different scopes: *node-level* techniques and *distributed* techniques. While there are many solutions trying to make a distributed storage system energy-proportional, many other researchers focus on making disks energy-efficient at the node level as a single component through algorithmic approaches – which is the scope of this paper. Dynamic power management (DPM) algorithms are the most popular technique to target energy saving in this scope. Basic DPM techniques attempt to transition disks to lower-power mode while experiencing relatively long idle periods However, without a careful design, they usually provide little energy saving or suffer from severe performance degradation due to non-negligible extra response time and energy costs imposed by spinning disks up and down when stand-by disks should service requests. Intuitively, the core idea of an effective DPM algorithm is to prolong the idling period of disks in order to allow them sleep longer in the lower-power mode and thus, boost power saving opportunity. We classify algorithmic techniques extending disks idleness period into two different categories as follows:

(1) The major approach taken in effective DPM algorithms is *data consolidation* or *data packing* that concentrates the frequently accessed data (hot fragments) into fewer number of disks (*hot disks*) in order to help other disks stay in idle mode longer. Thus, they usually perform corresponding inter-disk data migration in order to achieve the hot data consolidation goal. Popular Data Concentration (PDC) presented in [18] utilizes the load consolidation idea. However, it is not clear how their scheme adapts to different types of workloads as access frequency of data items are assumed static in this work while popularity of data fragments can significantly change with respect to time.

Another proposal of this category is presented in [22]. An efficient algorithm named dynamic block exchange (BLEX) is introduced that dynamically achieves load consolidation and performs necessary block exchange between disks. To the best of our knowledge, BLEX is the most effective algorithm in literature that tries to handle the dynamic I/O traces. Therefore, we will frequently make comparisons to BLEX in describing our solutions in the remainder of this paper. Our experiments will also use BLEX as the baseline.

Other similar proposals exploiting data packing are found in [19], [20], [21]. MAID [19] and PARAID [20] both directed their research efforts at the system level for energy conservation. Solutions presented in [19, 20, 21], in order to achieve data concentration goal, use replication (instead of data migration) that usually incurs significant space overhead. Also, note that RAID-based storage system is not the scope of our paper;
(2) The second approach to extend disk inactivity period is to manage I/O intervals via power-aware caching and prefetching algorithms. They extend the cache data replacement algorithm or prefetching techniques with energy-aware policies in order to gather hot data in cache while usually maintaining cold data on hard disks [10], [13], [14], [17]. In [14], Zhu presents power-aware cache replacement algorithm by defining energy cost for each block disposal from cache. Other existing power-aware solutions that focus on caching and prefetching rely on multi-tiered caches [11 ,12, 15]. They are based on using hybrid HDD and SSD drives as caching storage media. The drawback of the solutions in this category is the cost overhead such as the cost for SSDs or implementation cost for algorithmic solutions.

Moreover, there is another type of work that is similar to the second category in sense of employing cache to control I/O intervals; however, it utilizes redundancy as well with cache usage at the same time. It redirects some I/O requests to be served by redundant data available in cache or redundant disks. This scheme efficiently adapts to RAID storage layouts. [4], [16] use this approach for energy saving in RAID-based storage systems;

In addition to the two aforementioned categories, some other methodologies integrate DPM algorithmic techniques with hardware facilities to provide more power saving opportunities. One clear-cut instance is to take advantage of different rotation speeds available in multi-speed disks. Hibernator [24] and DRPM [25] dynamically configure disk speed based on the observed workload. Since DRPM is implemented within the disk controller and dynamic speeds are often not accessible to the operating system, it has restricted applicability. Note that, despite much research efforts, due to cost and physical constraints, multi-speed disk drives so far are not being commercialized in quantity [26] and thus will not be the focus of this paper.

Furthermore, a research work conducted under the EU GAMES project [27] proposed an adaptive mechanism for energy-aware data storage control [28]. It focuses on application-driven storage control in file storage—files are data units—for energy efficiency and thus, will not be the focus of our work. Also, another research paper utilized the application level I/O behaviors for power saving in storage systems cooperated with data-intensive applications [29].

As mentioned earlier, in the scope of distributed systems, there are many research efforts aiming at energy-proportionality in distributed storage systems [30, 31, 32, 33, 34, 35, 36, 37, 38, 39, 40, 41, 42]. The traditional data placement strategies existing in distributed systems restrict achievement of power proportionality [41] . Thus, all techniques in this category try to achieve energy proportionality either by extending the existing data placement strategies or by proposing novel data layouts. We have provided a more thorough survey on the related work in [43] that includes more details regarding the power-aware techniques discussed in this section.

In our previous work [23], an integrated DPM optimization model extended with MPC strategy is introduced that dynamically determines state (power mode) adjustments and plans for inter-disk fragment migration to achieve optimal tradeoff between power consumption and the query response time.

# 3. PROPOSED DPM OPTIMIZATION MODEL

In this section, we show the design of a DPM optimization model towards balance between energy consumption and performance. It is well-known that the arrival rate of data requests changes significantly in respect to time in I/O traces of database servers. This is particularly true in scientific database servers and OLTP servers. The SSDS SkyServer is a famous scientific database server that clearly shows significant changes in the server traffic rate [44]. The fluctuations in the data request arrival rate in SSDS server are observed to be up to 65% of the total arrival rate range. Also, [24] shows arrival rate changes in an OLTP trace that demonstrates remarkable arrival rate changes with respect to time. The major problem of previous contributions is that they cannot efficiently enough adapt to dynamic I/O workloads. We solve this issue by integrating model predictive control (MPC) strategy in an optimization model to enable optimization actions in an online fashion. Section 3.4 describes in detail how our optimization model integrates the MPC technique in order to capture the dynamic changes in data access frequency. In addition to the MPC strategy, another advantage of the DPM model is that it explicitly includes fixed penalty cost on disk status change to avoid excessive spin up and down operations that have expensive response time and energy costs. However, this is rather considered subjectively in BLEX algorithm.

Given the arrival rate changes in dynamic I/O workloads, we partition the planning horizon into multiple periods where the arrival rate in each period can be modeled by a constant. We formulate a model as a (nonlinear) mixed integer program (shown in Section 3.1) where the objective function is the overall cost from all energy consumption elements in the storage system during one epoch. At the beginning of each epoch, based on the observed workload and the predicted workload for the epoch, the model configures the optimal disk state setting and corresponding inter-disk data migration such that the energy consumption (aforementioned objective function) during the epoch is minimized while maintaining query response time quality. In order to avoid the disk overloading problem, the model performs load balancing between the overloaded disk(s) and other active disks at the beginning of each epoch.

The length of the epoch should be short enough to capture changing arrival rates and also long enough to accommodate disks transition cost and data migration periods as well as tolerable number of on/off actions on disks in order to not damage their lifetime services. Considering arrival rate change patterns existing in database I/O traces, we verified different epoch length values to determine an efficient value that fulfills the above requirements. Based on our sensitivity analysis described in Section 5.7, the energy saving ratio is insensitive to the epoch lengths larger than 30 minutes. Therefore, we determined the epoch length to be 30-minute long since it captures arrival rate changes effectively while exploiting energy savings.

The assumptions in the model are as follows:

(1) Any fragment can only be migrated once in a period;
(2) Power rate of each disk depends on its current state (rotation speed);
(3) The data migration cost (time) between any two disks is intuitively proportional to the total size of data fragments to be migrated and is independent of the source and destination disks;
Table 1 introduces the main parameters and indices used in the model development. Table 2 introduces the list of decisions variables used in our DPM optimization model including binary, integer and continuous variables.

Table 1. DPM Optimization Model Parameters

| Name | Description |
|---|---|
| $i$ | Index of disks, $i = 1, \ldots, I$ |
| $j$ | Type of data fragmentation based on request arrival rate pattern – data belonging to each type share the similar request arrival rate pattern given data correlations in queries , $j = 1, \ldots, J$ |
| $\lambda_{j,t}$ | Hotness level of fragment type $j$ in period $t$ – hotness level is determined by the observed data request arrival rate |
| $k$ | State of disk |
| $Sc_i$ | Storage capacity of disk $i$ |
| $c_j$ | Migration cost of fragment type $j$ |
| $b_j$ | Block size of fragment type $j$ |
| $ed_i$ | Energy to spin down disk $i$ |
| $ep_i$ | Energy to spin up disk $i$ |
| $p_{i,k}^t$ | power consumption of disk $i$ at $k$ spinning state in period $t$ |
| $\Gamma$ | Response time penalty parameter |
| $maxfrag$ | Disk maximum number of data fragments |
| $\lambda^{max}$ | Maximum data fragment hotness level (arrival rate) |
| $M$ | Maximum no. of blocks in a disk |

Table 2. Decision Variables

| Name | Type and Description |
|---|---|
| $x_{i,j}^t$ | Integer - Quantity of $j$ type fragment on disk $i$ in period $t$ |
| $y_{j,i_1,i_2}^t$ | Integer - Quantity of $j$ type fragments migrated from $i_1$ to $i_2$ at the end of period $t$ |
| $s_{i,k}^t$ | Binary - Equals to 1 if disk $i$ is in state $k$ in period $t$ |
| $u_i^t$ | Binary - Equals to 1 if disk $i$ should be spun up in period $t$ |
| $d_i^t$ | Binary - Equals to 1 if disk $i$ should be spun down in period $t$ |
| $T_i^k$ | Continuous - Response time penalty of disk $i$ in state $k$ |

## 3.1. FORMULATION OF DPM OPTIMIZATION FOR MULTI-STATE DISKS

Our objective is to minimize the energy consumption within each epoch period. The total energy consumption during an epoch consists of four elements. The first part is the primary energy consumption of the disk storage that depends on disk states (rotation speed) and number of disks spinning in each state. It is independent of the migration operations. The second part is the energy consumed during the migration time which strictly depends on the total fragment size of migration. And, the rest of energy consumption includes energy costs for disk spin-up and spin-down operations. The objective function is shown in the following equation:

$$min \sum_{t=1}^{T}\sum_{i=1}^{I}\sum_{k} p_{i,k}^t \ s_{i,k}^t + \sum_{t=1}^{T}\sum_{j=1}^{J}\sum_{i_1=1}^{I}\sum_{(i_1 \in I, i_2 \neq i_1)} c_j \ y_{j,i_1,i_2}^t + \sum_{t=1}^{T}\sum_{i=1}^{I} ep_i \ u_i^t + \sum_{t=1}^{T}\sum_{i=1}^{I} ed_i \ d_i^t + \sum_{t=1}^{T}\sum_{i=1}^{I}\sum_{k} \Gamma \cdot s_{i,k}^t \ T_i^k$$

(1)

The physical and logical constraints in the model are as follows:

(1) Size of fragments stored in a disk can never exceed the disk capacity;
(2) Disks must stay in a certain state during an epoch period;
(3) During any epoch $t$ , there must be at least one active disk serving the data requests;
(4) Any fragment can only migrate once in a certain epoch $t$ ;
(5) A disk in stand-by mode is not considered as source or destination for data migration;

(6) There is a limit for data migration time (H) that represents the data transfer limit for any disk within an epoch. The migration limit by default is set to half of the epoch;

The following equations represent the aforementioned constraints respectively:

$$\sum_{j=1}^{J} b_j \; x_{i,j}^t \leq Sc_i \quad \forall i, t \tag{2}$$

$$\sum_{k=1}^{K} s_{i,k}^t = 1 \quad \forall i, t \tag{3}$$

$$\sum_{i=1}^{I} s_{i,k=1}^t \leq I - 1 \quad \forall t \tag{4}$$

$$\sum_{i_2} y_{j,i,i_2}^t \leq x_{i,j}^t \quad \forall i, t, j \; (i_2 \neq i) \tag{5}$$

$$y_{j,i_1,i_2}^t \leq M \cdot \sum_{k=2}^{K} s_{i_1,k}^t \quad \forall i, t, j \; (i_2 \neq i_1)$$

$$y_{j,i_1,i_2}^t \leq M \cdot \sum_{k=2}^{K} s_{i_2,k}^t \quad \forall i, t, j \; (i_2 \neq i_1) \tag{6}$$

$$\sum_j \left( \sum_{i_2} y_{j,i,i_2}^t + \sum_{i_2} y_{j,i_2,i}^t \right) \leq H \quad \forall i, t, j \; (i_2 \neq i) \tag{7}$$

Also, the migration equation that links $x_{i,j}^t$ and $y_{j,i_1,i_2}^t$ is:

$$x_{i,j}^t + \sum_{i_1 \in I, i_1 \neq i} y_{j,i_1,i}^t = x_{i,j}^{t+1} + \sum_{i_2 \in I, i_2 \neq i} y_{j,i,i_2}^t \quad \forall i, t \geq 1, j \tag{8}$$

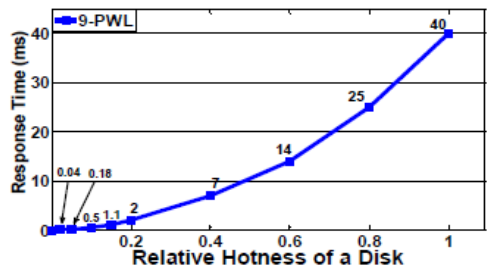

Figure. 2. 9-PWL Function of Response Time Model

And, in order to determine the binary indicating variables related to spin up and down of disks, the following equations are used in the model:

$$\sum_k k \; s_{i,k}^t - \sum_k k \; s_{i,k}^{t+1} \leq u_i^t \tag{9}$$

$$\sum_k k \; s_{i,k}^{t+1} - \sum_k k \; s_{i,k}^{t} \le d_i^t \tag{10}$$

## 3.2. TWO-STATE OPTIMIZATION MODEL

We develop DPM optimization model for two-state disk storage system. It is easy to obtain the model formulation for two-mode (active and stand-by) disk storage by setting two values for parameter $k$ (1 or 2) in the general formulation provided in the previous section for multi-mode disk storage. The equations related to two-mode optimization model are provided in [45]. The general DPM optimization model assumes 10 levels of data popularity (hotness level) for data fragments based on the observed data request arrival rate. We believe that having 10 levels is sufficient to accurately classify data blocks based on the hotness level (if more resolution would be needed, the model can certainly have more levels that indeed reduce the MPC computational time). An important feature of two-state optimization model is that the least and the second least popular data stay in original disks. This will help to minimize the migration cost.

## 3.3. RESPONSE TIME MODELING

The expected response time of a disk is a function of its spinning state and the total data arrival rate. Thus, if we consider the state of disk constant, the response time of the disk is a convex function with respect to its relative hotness level (defined in the following equation) with increasing first derivative order. We modeled this function by using piecewise linear (PWL) functions in our optimization model since they are widely used to approximate any arbitrary function (especially convex functions) with high accuracy. The input of PWL function is relative hotness of a disk. The relative hotness of a disk is calculated by following equation:

$$\lambda_{i,t} = \frac{\sum_j \lambda_{j,t} \; x_{i,j}^t}{\lambda^{max} \; maxfrag} \tag{11}$$

where $\lambda_{i,t}$ is the relative hotness level of disk $i$ in period $t$ and $0 \le \lambda_{t,i} \le 1$, $1 \le \lambda_{j,t} \le 10$ is the popularity of fragment type $j$ in period $t$, $maxfrag$ is maximum number of fragments in a disk and $\lambda^{max}$ is upper bound for data popularity (hotness level). We define $L$ as the number of linear functions to approximate the response time. It is well known that PWL functions can represent arbitrary functions to any accuracy by simply increasing the number of segments ($L$) to the point of desired accuracy. Thus, we verified different $L$ values for approximation of the response time convex function. We decided to use 9-PWL function shown in Figure 2 for two-state disk storage system since it approximates the convex function with high accuracy.

## 3.4. MODEL PREDICTIVE CONTROL (MPC)

The presented optimization model is rather static while our actual system works in a dynamic on-line environment. Therefore, we extend the model to accommodate dynamic scenarios by using model predictive control (MPC) technique to solve this issue. MPC, also known as receding horizon control (RHC) or rolling horizon control, is a form of control strategy to integrate optimization. Specifically, the current control action is obtained in an on-line fashion where, at each sampling instant, a finite horizon optimization problem (which is (1)-(10) in our context) is solved and its optimal solution in the first stage is applied as the current control decision while remaining solutions will be disregarded. Such procedure repeats along the whole control process. Therefore, all controllable variables (such as disk status and response time) for the first period are implemented in the MPC. It has been observed that MPC is a very effective control strategy with reasonable computational overhead [46].

The prediction information on workload arrival rate is provided to the MPC optimization model. This information plays a key role in developing an accurate underlying mixed integer program for the DPM model since any mis-prediction of data request arrival rates could cause the model to produce a solution with a less desired quality. However, as observed in many other applications of MPC, since only the first stage solution will be implemented and remaining parts will be ignored, MPC control strategy is robust to poor predictions and has a strong adjustment capability [47]. Also, our experimental results in Section 5.6 clearly demonstrate strong robustness of MPC integrated in the optimization model against mis-predictions.

### 3.5. SOLVING STRATEGY

Our initial attempt to find solutions to the two-state model is to implement and solve the model in the well-known CPLEX solver. The solver is installed on a server which is connected to another server running the widely used disk simulator, *DiskSim* [48], which is utilized as an accurate and reliable simulation platform by many related works. In other words, the model solution is integrated in the disk storage system simulated in DiskSim. Technical details regarding the experimental simulations are provided in Section 5.

## 4. PROPOSED HEURISTIC DPM ALGORITHM

In this section, we propose a fast heuristic DPM algorithm that efficiently determines disk power-mode states and inter-disk data migration based on the observed I/O workload in order to maximize power saving while satisfying the response time bound. The algorithm is designed for two-state disk storage systems since, as mentioned earlier, multi-mode disks so far have not been widely commercialized.

MPC is known for its overhead when the scale of the problem grows large [28]. Thus, the MPC integrated in the DPM optimization model incurs significant overhead for finding its optimal solution at each epoch in large-scale disk storage systems. There are some discussions on how to reduce the MPC overhead (i.e. via hardware implementation) presented in [49] that are not applicable to the research in this paper. Therefore, via analyzing the optimization model to identify structural properties of the optimal solution, we propose our heuristic DPM algorithm that provides near optimal power saving solution (near to that of optimization model) with fast computational time. As shown in Section 5, the heuristic algorithm significantly outperforms the BLEX in terms of energy saving and response time.

As shown in the following section, our heuristic algorithm supports *merge* and *split* operation between disks (in terms of their data content) and performs analytical comparison for decision on each operation. In merge operation, the analytical comparison is considered between two cases: merging active disks to one disk; and keeping the current status of disks. It performs the similar comparison for split operation on disks. Such analytical comparisons definitely lead to more efficient energy management in disk storage—which accounts for an advantage of the heuristic algorithm over BLEX algorithm.

### 4.1. ASSUMPTIONS

We use the same epoch concept (and value) described in the previous section for the heuristic algorithm implementation. Furthermore, given that transition time between disk states (usually several seconds) is much smaller than epoch length (30 minutes), we can ignore the energy consumption caused by disk state change comparing to that consumed during an epoch.

Therefore, it is assumed that disk status change will not incur energy consumption/cost. Also, it is assumed that data migration will not incur extra power consumption as it only imposes extra penalty cost on response time due to fixed data migration time—which is based on the total size of the transferred data.

## 4.2. PROPERTIES OF THE ALGORITHM

We define a cost function for a disk called $f$ . The properties of function $f$ are as following:

(a)  $f$ is defined for a single disk and it consists of two parts: fixed energy cost of an active disk during an epoch, say *EnergyCost*  (equals to *0* if it is in stand-by mode); and the response time penalty cost that depends on disk hotness level. Note that the hotness level of the disk means the weighted summation of its data hotness levels—similar to (11) in Section 3.3;
(b) It monotonically increases with respect to disk hotness level;
(c) $f$  is convex function with respect to the hotness level of disk; We use the 9 piece-wise linear function described in Section 3.3  to represent the convex function for the response time penalty part of function $f$ .

The proposed heuristic algorithm has the following features:

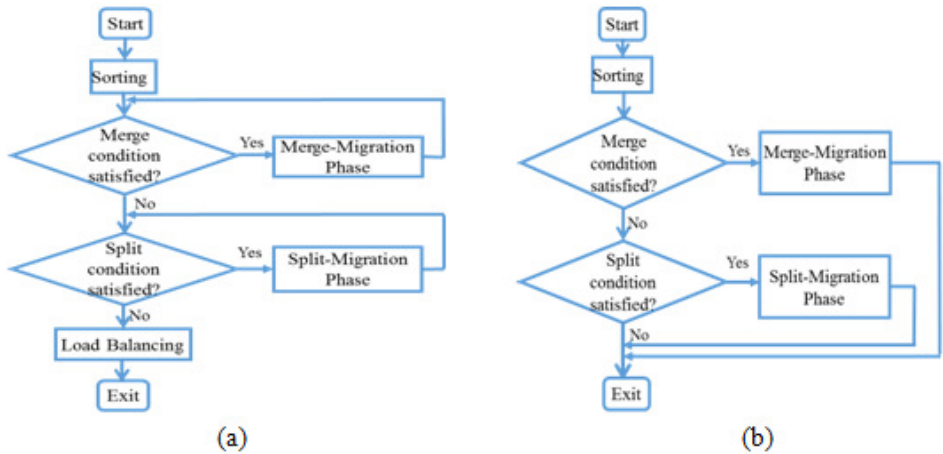

Figure 3. Flowcharts For The Sequential Pairing Algorithms: (a) One-To-One Heuristic Algorithm; (b) One-To-Many Heuristic Algorithm

(1) The best data migration plan happens when the data is equally distributed on active disks. Otherwise, it can be proven that the objective function value is not optimal;
(2) Inter-disk data migration is performed from disk(s) supposed to be cold to disk(s) supposed to be hot so that the data migration cost is minimized;
(3) Once a disk is active, its energy consumption only depends on its current state and independent of hotness level of the disk;

## 4.3. ALGORITHM DESCRIPTION

In this section, we present the fast heuristic DPM algorithm which is called the Sequential Pairing algorithm.  First, we define two types of possible data transfer between source and destination disks. Then, we present two versions of the sequential paring algorithm corresponding to each type of data transfer assumed for fragment migration. We consider two different possible types of data transfer between disks as follows: one-to-one transfer in which data migration can be performed only between two disks; One-to-many (or many-to-one) transfer that allows data migration from one disk to multiple disks and vice versa.

### 4.3.1. ONE-TO-ONE SEQUENTIAL PAIRING ALGORITHM

The one-to-one sequential paring algorithm in general suggests two different types of data migration: *merge-migration* that merges all data of an active disk into another active one in order to turn off the source disk; and *split-migration* that splits the data of an active disk evenly with another inactive disk in order to turn on the destination. Figure 3-(a) shows a flowchart that describes the high-level flow between different steps of this algorithm. The one-to-one sequential algorithm is as follows:

*Step 1.* Sort disks according to their hotness level in ascending order. Let $L$ be the sorted list.
*Step 2. Merge-migration phase:* Pick the first sequential pair from the beginning of $L$ (if there is no available pair, continue to step 3).

> *2.1.* If the cost function is decreased by merging the two disks, perform the merge-migration and update the sorted list $L$ with new hotness levels of disks. Repeat step 2.
> *2.2* Otherwise, there is no merge-migration needed. Go to step3.

*Step 3. Split-migration Phase:* If there is any inactive disk, pair it with the disk having the maximum hotness level in sorted list $L$ . Otherwise, go to step 4.

> *3.1.* If the cost function is decreased by splitting the data between the pair of disks, activate the stand by disk and perform split-migration. Update the sorted list $L$ with new hotness level values of disks and Repeat step 3.
> *3.2.* Otherwise, there is no split-migration needed and continue to step 4.

*Step 4. Load-balancing Phase:* Evenly distribute the data between active disks.

### 4.3.2. ONE-TO-MANY SEQUENTIAL PAIRING ALGORITHM

Similar to one-to-one version of the algorithm, one-to-many sequential pairing algorithm also assumes two types of migrations: *merge* and *split*. In merge-migration, it migrates all data from one source disk *evenly* to many destination disks in order to turn off the source disk. And, in split-migration, it allows migration of data from multiple source disks to one destination disk in order to activate the destination disk (many-to-one). Also, similar to the previous version of the algorithm, one-to-many sequential pairing mainly consists of two phases: merge-migration and split-migration. Under the assumption of fragment migration to/from multiple disks, we can achieve the ideal situation in both merge and split phases. In both phases, we compute the ideal data allocation plan that provides the optimal number of active disks, say $n^*$, and the average hotness level, say $\bar{v}$, over $n^*$ disks. The ideal data allocation plan can be easily obtained from following equation:

$$\bar{v} = \frac{\sum_{j=1}^{n} v_j}{n^*} \tag{12}$$

where $n$ is the total number of current active disks and $v$ represents the hotness level for each disk. Next, we perform required migrations (depending on the phase) to reach average hotness level $\bar{v}$ on all $n^*$ active disks. Figure 3-(b) shows a flowchart that demonstrates the high-level flow between different steps of this algorithm. The one-to-many sequential pairing algorithm is described in detail in the following:

*Step 1.* Sort disks according to their hotness level in ascending order. Let us assume $L$ is the sorted list and $n$ is the number of current active disks.

*Step 2. Merge-migration Phase:*

    <u>*2.1*</u> *Optimal data allocation plan:* decrement $n$ by one and compare the new cost function with the current cost function. If the cost function is not reduced, go to step 3. Otherwise, repeat decrementing $n$ one-by-one as long as the cost function is reduced in each decrement. $n^*$ is determined $n - k$ where $k$ is the number of past successful decrements by which the cost function was reduced. Calculate $\bar{v}$ by using (12).

    <u>*2.2*</u> *Merge-migration plan:* The data on the first $k$ disks in the list $L$ is merged to the rest $n^*$ disks, say *optimal list*, through one-to-many merge migrations. The detailed routine performing the merge-migration plan is provided in [45].

    <u>*2.3*</u> *Load Balancing:* Evenly distribute the data among the disks in the optimal list such that hotness level $\bar{v}$ is reached for each disk. Then, exit the algorithm.

*Step 3. Split-migration Phase:*

    <u>*3.1*</u> *Optimal data allocation plan:* increment $n$ by one and compare the new cost function with the current cost function of $n$ disks. If the cost function is not reduced, exit the algorithm. Otherwise, repeat incrementing $n$ one by one as long as the cost function is reduced in each increment. $n^*$ is set to $n + k$ where $k$ is the number of successful increments by which the cost function was reduced continuously. Calculate the optimal data allocation plan ($\bar{v}$) by using formula given in (12).

    <u>*3.2*</u> *Split-migration plan:* Add $k$ newly activated disks to the list $L$. Perform many-to-one split migrations to achieve $\bar{v}$ on each newly activated disk. The detailed routine that performs split-migration plan is provided in [45].

    <u>*3.3*</u> *Load Balancing:* Evenly distribute data among the original $n$ disks such that hotness level $\bar{v}$ is also reached on each of these disks. Then, exit the algorithm.

As shown in next section, the proposed heuristic algorithms adapt to large-scale disk storage systems effectively since they provide fast and efficient solution in terms of both energy saving and response time that are near to optimal solution provided by the optimization model.

# 5. EMPIRICAL EVALUATION

We have conducted extensive set of simulations using broad range of I/O workloads to validate the effectiveness of our proposed techniques. We have compared our solutions in terms of energy saving ratio and average response time with those of BLEX algorithm.

## 5.1. SIMULATED DISK STORAGE SYSTEM

The disk storage used in our simulations consists of an array of conventional hard disks. Each disk is configured as in independent unit of storage. We have simulated the array of 15 disks in *DiskSim*, a widely used disk storage simulator by many related research works. The hard disk model used in simulations is IBM Ultrastar 7K6000 [9]. The main specifications of this hard disk are summarized in Table 3.

Table 3. Hard Disk Specifications

| Description | Value |
|---|---|
| Disk model | Ultrastar 7K6000 |
| Standard Interface | SAS |
| Rotational Speed (active) | 7200 rpm |
| Rotational Speed (stand by) | 3600 rpm |
| Disk Capacity Size | 2 TB |
| Seek time (average) | 8 milliseconds |
| Power in active mode | 11 W |
| Power in idle mode | 7.7 W |
| Power in sleep mode | 1 W |
| Spin up time | 15 Seconds |
| Spin down time | Immediately |
| Transfer Rate (MB/Sec) | 202 |

## 5.2. SYNTHETIC WORKLOAD GENERATOR

We developed a workload generator written in C to synthesize disk access I/O workloads based on popular database TPC benchmarks. We follow the well-known $b/c$ model in generating a workload of a series of random data read operations ($b$% of all read operations is against $c$% of the data) [50]. It is well known that database tuple access pattern is highly skewed and can be described as an 80/20 or even a 90/10 model [51]. Zipf probability distribution is used in the generator to produce $b/c$ model. The default $b/c$ model used in simulations is set to 80/20. We have used Gamma distributions in our workload generator to reflect the dynamic behaviour of database data access frequency. Given the data correlations among database tuples in queries, the access frequency change pattern of each data fragment type is represented by a Gamma distribution.

## 5.3. EXPERIMENTAL PLATFORM

Our model is integrated in the disk simulator. The BLEX algorithm is also implemented in the DiskSim, as the comparison target, based on its description in [22]. We enhanced DiskSim with a multi-speed disk power model where the disk power consumption rate is proportional to disk rotation speed. Also, it is augmented with extra features such as dynamic disk spin up (and down), disk state adjustment and inter-disk data migration.

The predicted access frequency (hotness level) for each fragment type for the next $k$ epochs is provided to the model along with the observed fragment type access frequencies in the previous epoch. The prediction is performed by the prediction and autoregressive modeling methods in Matlab. In particular, an autoregressive model is developed based on the observed data access frequency. Then, the prediction method forecasts fragments access frequency for the next $k$ epochs ahead based on the identified model and the observed data frequency.

## 5.4. SIMULATION RESULTS AND COMPARISONS

In this section, we describe the experimental results in terms of energy saving ratio (normalized with no power saving scheme applied) and average response time using extensive set of synthetic traces. The workloads are classified in two categories: *dynamic* and *static*. The data request arrival rate changes in respect to time in dynamic workloads while it is static over time in static traces.

### 5.4.1. DYNAMIC I/O WORKLOAD TRACES

We compare the performance of DPM optimization model and heuristic algorithms with BLEX algorithm under comprehensive set of dynamic traces whose data request arrival rate changes

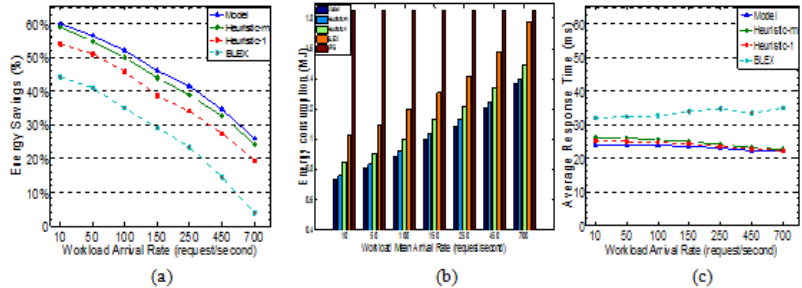

Figure 4. Experimental results under dynamic I/O traces with different mean data arrival rates: (a) Energy saving results; (b) Total power consumption of the disk storage system using different power saving schemes; (c) Average response time results

over time. As mentioned earlier, we believe that BLEX algorithm cannot strongly enough handle the dynamic aspect of I/O. The reason is that it performs inadequate data migration and keeps some data in stand-by disks and therefore, to be able to adapt to dynamic changes in data request arrival rates, it might pay some penalty related to spinning stand-by disks up and down. On the other hand, MPC control scheme integrated in the optimization model effectively captures the dynamic behavior of I/O workload traces over the time. The following experimental results demonstrate this characteristic of MPC control strategy which is the key advantage of our DPM optimization model over previous proposals.

**Energy Saving Results.** Figure 4-(a) shows energy saving ratio for various dynamic I/O traces with different mean data arrival rates (request/second) under all power saving schemes. Results in Figure 4-(a) clearly show that the DPM optimization model and the heuristic algorithms significantly outperform the BLEX algorithm. Optimization model, Heuristic-m (one-to-many) and Heurisitic-1 (one-to-one) save energy up to 60%, 58% and 54% respectively. The MPC-based optimization model outperforms BLEX with difference of at least 16% and up to 23% in energy savings. It saves 19% more energy on average than BLEX.

According to the results in Figure 4-(a), one-to-many sequential pairing algorithm provides power saving solutions near to that of the optimization model. It significantly outperforms BLEX algorithm with difference of at least 14% and up to 21% in energy savings. Heuristic-m saves around 16% more energy on average than BLEX algorithm. One-to-one sequential pairing algorithm also provides better power saving results than that of BLEX. In comparison with BLEX, It provides 11% improvement on energy saving on average based on the results.

Figure 4-(b) shows the total power consumption of the disk storage system for each power saving method compared to that of no power saving (NPS) method applied to disk storage, where all disks constantly run in active mode (shown as a red bar in the figure). We can conclude that DPM optimization model is dominant in power saving. One-to-many sequential algorithm outperforms the other version as well as BLEX algorithm in reducing the power consumption. It provides energy saving near to that of the optimization model. One reason is that it always attempts to evenly relocate data among ideal number of active disks. On the other hand, as it is shown in next section, it has an overhead, however acceptable, on average response time due to additional inter-disk data migration towards load balancing.

**Average Response Time Results.** Saving power potentially incurs increase in query response time. Thus, it is important to measure the response time effected by power saving schemes to ensure that high quality of service for queries is still maintained. This will help us in understanding the limitations of our model and algorithms.

Figure 4-(c) shows the average I/O response time for all power saving schemes under several workloads with different mean arrival rates. Note that the computational time to obtain the solution for all power saving schemes is up to a second, which is apparently ignorable compared to the epoch length (30 minutes). Thus, it is excluded from the average response time computations. The results show that optimization model provides better response time than all other schemes. One reason, other than response time considerations in its optimal power-performance tradeoff, is that it takes into account the predicted information on data access frequency in next epochs for its solutions. Both versions of heuristic algorithm provide response time results near to that of the optimization model. BLEX has longer response time than other schemes. The reason is that it performs little data migration since it only migrates blocks in cold disks to hot disk when the blocks in cold disks are accessed.

## 5.4.2. STATIC I/O WORKLOAD TRACES

In this section, we describe the experimental results in terms of power saving ratio and average response time for all power saving schemes under extensive set of I/O workload traces whose arrival rate is *static* over the time.

**Energy Saving Results**. Figure 5-(a) demonstrates the power saving results for DPM optimization problem, heuristic algorithms and BLEX algorithm. The energy saving is normalized with the energy consumption of the storage system where all disks are always active with no power saving method applied. Energy saving ratio is shown for several I/O traces with different static data arrival rates (request/second). Based on the results shown in Figure 5-(a), DPM optimization model is dominant in saving energy up to 72%. The optimization model, even under static traces, saves up to 8% more energy than BLEX. The heuristic algorithms also shows high performance near to the optimization model in terms of power saving. Specifically, Heuristic-m algorithm closely follows optimization model in saving energy.

**Average Response Time Results**. Figure 5-(b) depicts the average response time results for all schemes under static arrival rates. Note that Figure 5-(b), similar to response time results in previous section, excludes the computational time taken to compute solutions for each power saving method. It shows that optimization model provides the best performance comparing to all other methods. The heuristic algorithms also demonstrate reasonable response time close to that of optimization model. Both versions of the heuristic algorithm outperform the BLEX algorithm in terms of average response time. According to the experimental results shown in Figure 5, although our model is designed to handle dynamics, it is also general to cover scenarios with static workloads.

## 5.5. LARGE-SCALE DISK STORAGE SIMULATION

We conducted experimental simulations using extensive set of *dynamic* I/O workloads to evaluate the performance of our proposed ideas, especially the heuristic algorithm, in large-scale disk storage systems. Thus, we have extended the aforementioned experimental platform to simulate a 100-disk storage system. Since there is larger number of disks involved in inter-disk data migrations at the beginning of each epoch, it incurs longer time for the entire disk storage to adjust to required configurations. Therefore, we have extended the epoch length to 60 minutes to amortize the longer inter-disk data migration periods in the simulated large-scale storage system. Our experimental results showed that the computational time for finding the optimal solution by the DPM optimization model in the large-scale storage system takes on average 20-30 minutes as expected due to the MPC overhead, as discussed earlier. However, both heuristic algorithms demonstrate fast computational time less than a few seconds that can be apparently ignorable compared to the epoch length (60 minutes). The BLEX algorithm also incurs a few-second computational time similar to that of heuristic algorithms.

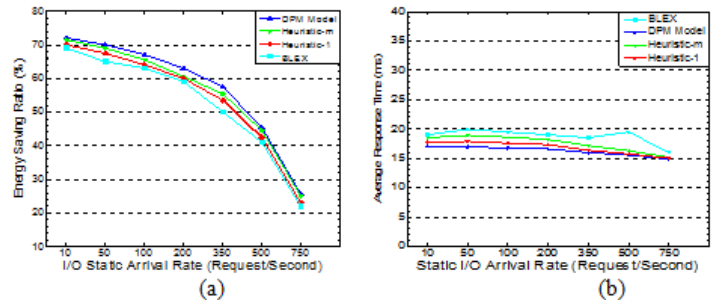

Figure 5. Experimental results under static I/O traces with different data arrival rates: (a) Energy saving results; (b) Average response time results

**Energy Saving Results.** Figure 6-(a) shows energy saving results related to all power saving methods under broad range of dynamic I/O traces with different mean arrival rates. The energy savings are normalized to the energy consumption of disk storage where no power saving method is deployed. It clearly shows that the optimization model achieves the best power saving result, however, with long computational time. According to the experimental results, one-to-many sequential pairing algorithm provides power saving solution near to optimal solution provided by DPM optimization model and saves only 4 % less energy on average than optimization model, however, with fast computational time. It also significantly outperforms the BLEX algorithm with around 16% more energy saving on average. Heuristic-1 algorithm also provides higher energy saving ratio than BLEX and shows 9% improvement on energy saving on average compared to BLEX. Based on the experimental results, the proposed heuristic algorithms, specifically Heuristic-m, demonstrate fast and efficient power saving solutions for large-scale storage systems.

**Average Response Time Results.** We validate the effect of power saving methods on the average response time results in the 100-disk simulated storage system. Note that the average response time results shown in Figure 6-(b) exclude the computational time taken to compute the solutions for each power saving method. The quantitative results for the computational time are separately reported in Figure 6-(c). It depicts the measured computational time for both versions of the heuristic algorithm as well as the BLEX algorithm for I/O workload traces used in this experiment. The measurements are in terms of the average computational time of each algorithm during an I/O trace. Figure 6-(b) clearly shows that both heuristic algorithm provide significantly better response time than that of BLEX algorithm except only for one trace with arrival rate 100 request/sec in which BLEX has slightly better response time than Heuristic-m and slightly worse than Heurisitic-1. Heuristic-m and Heuristic-1 achieve around 10 and 9 milliseconds faster response time on average than that of BLEX algorithm respectively.

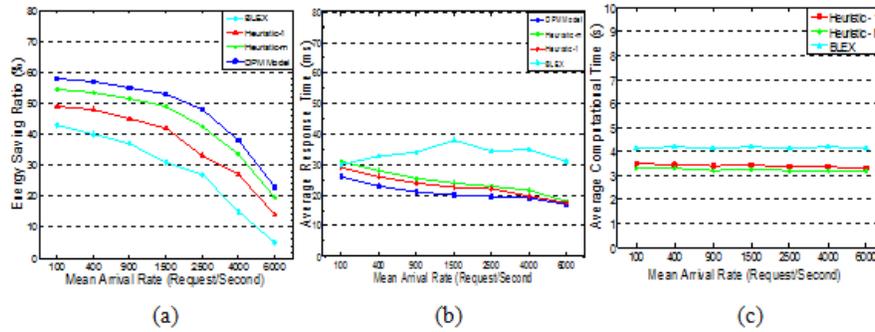

Figure 6. Experimental results for large-scale storage system under dynamic I/O traces with different mean arrival rates: (a) Energy saving results; (b) Average response time results; (c) Average computational time results;

Based on the experimental results in this section, the proposed heuristic DPM algorithm demonstrates near optimal power saving solutions (near to that of MPC-based optimization model) in both energy saving and average response time with fast computational time. Thus, we conclude that the heuristic algorithm, especially one-to-many sequential pairing, can be integrated in large-scale disk storage systems, where finding the optimal solution might be long, to achieve efficient energy saving solutions within short periods of computational time.

## 5.6. MPC ROBUSTNESS AGAINST MIS-PREDICTIONS

In this section, via running extensive set of experiments in the systematic way, we evaluate the robustness of MPC technique integrated in the optimization model against mis-predictions in data request arrival rates. Note that the simulated disk storage used for experiments in this section is the 15-disk storage described in Section 5.1. As discussed in section 3.4, the prediction information on workload arrival rate that is provided to the MPC-based optimization model plays an important role in accurate development of the model. Therefore, any mis-prediction in data request arrival rate could cause the model to produce a solution with less desired quality. However, MPC strategy is robust to poor predictions and has a strong adjustment capability. As mentioned earlier in Section 3.4, only the first stage solution will be implemented and the remaining parts will be ignored. We evaluate the aforementioned feature of MPC by running extensive set of experiments.

First, we randomly place noises (mis-predictions) in the data request arrival rate predictions for future epochs—which are fed into the optimization model as an input. Then, we gradually increase the percentage of noises in the predictions to monitor how the solution given by the optimization model deviates from the optimal solution (correct) where predictions are error free and accurate. Deviation from the error-free solution leads to a non-optimal trade-off between energy saving and response time. Note that the intensity of each single mis-prediction (difference from the correct prediction value) is randomly determined. The following plots in Figure 9 show the solution produced by the model against different amount of prediction errors in multiple experiments. Note that model solution on disk state configuration is represented as the number of disks to be transitioned to stand-by mode.

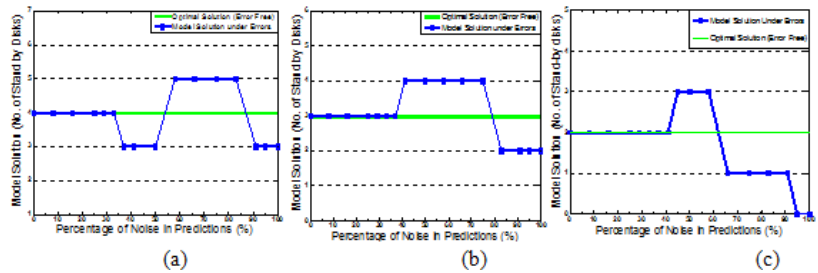

Figure 7. MPC control results under different prediction error rates

The results demonstrate robustness of the MPC strategy integrated in the optimization model. According to Figure 7-(a), 7-(b) and 7-(c), the model output under mis-predictions matches the error-free solution under error rates up to 36%, 37% and 41% respectively. For further error rates up to 93%, the model solution deviates from the optimal solution with the minimum possible difference in terms of disk configuration (one-disk difference) in all these experiments. It deviates with 2-disk difference in terms of disk configuration in only one experiment (Figure 7-(c)) where prediction error rate is greater than 95%. It is important to note that the trend of results seen in the experiments in Figure 7 was observed in all other experiments as well.

Second, we performed another set of experiments in a systematic way to verify the effect of prediction errors of data request arrival rate on the solution of the MPC-based DPM model. Given that the model receives the prediction information for multiple future epochs, it is intuitive to consider less confidence in predictions for farther epochs. In other words, prediction error rate increases for farther epochs in future. Therefore, we assume a fixed error ratio *limit* for predictions of each epoch and this error limit is relatively larger for farther epochs ahead. More specifically, we assume that error ratio limit increases linearly with a constant slope for data access predictions in farther epochs. Therefore, in this type of experiment, all predictions of data access frequency for all future epochs are imposed to errors, however, with different error bounds. Note that the distribution of error rates among all predictions related to a single epoch is even distribution based on the corresponding error bound for that epoch.

Predictions are generated for up to four epochs ahead in this type of experiment. The error ratio limits for future epochs are represented as an *error set*. As an instance, error set [20%-30%-45%-67.5%] represents four error ratio limits corresponding to predictions for the next four epochs in future. For example, based on this error set, prediction errors related to the first epoch in future can vary between -20% and +20% (from the error free prediction value) with even distribution among all predictions for this epoch. Similar to the previous experiment, we monitor the effect of mis-predictions on the quality of the solution produced by the model by applying a wide range of error sets to each experiment. The error sets vary from set [5%-7.5%-11.2%-16.9%] up to set [29.5%-44.2%-66.3%-99.5%] in each experiment.

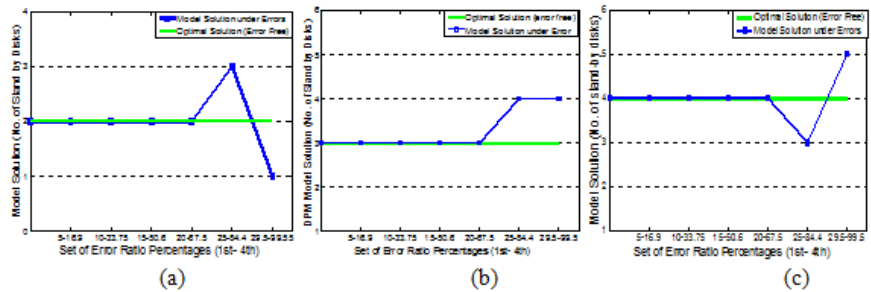

Figure 8. MPC control results under different prediction error rates

Figure 8 clearly shows strong robustness of MPC optimization model against poor predictions. The model solution under errors matches the optimal error-free solution for all error sets that are equal or less than the set $[20\% \text{-} 30\% \text{-} 45\% \text{-} 67.5\%]$. It deviates from optimal solution with the minimum possible difference in terms of disk configuration (one-disk difference) for all other error sets greater than set $[20\% \text{-} 30\% \text{-} 45\% \text{-} 67.5\%]$ in all experiments shown in Figure 8. Note that the same trend of results was observed in all other experiments of this type performed for MPC robustness verification. Based on the experimental results in this section, it can be concluded that the MPC strategy integrated in our DPM optimization model is strongly robust against poor predictions and has powerful adjustment capability. As mentioned in Section 3.4, the reason is that only the first stage solution will be implemented and remaining parts will be ignored.

## 5.7. Effect Of Epoch Length On Energy Saving

We explore the effect of the epoch length on the energy saving ratio by running extensive set of experiments using various number of database dynamic I/O workloads with different arrival rate change patterns. For each particular I/O workload, a large number of different epoch lengths are chosen and the corresponding energy saving ratio are measured separately. In order to synthesize the arrival rate change pattern of I/O traces, we have used Gamma distributions (parameters $k, \theta$) as described in Section 5.2. The epoch length should be long enough to accommodate the disk state adjustments and data migration and also it should be short enough to capture the dynamic data request arrival rate changes. Therefore, we intuitively introduce a reasonable lower and upper bound for the epoch length as 10 minutes and 240 minutes respectively. In each experiment, the epoch length is incremented by 15-30 minutes for a particular I/O trace and the corresponding energy saving ratio is measured separately in order to observe the effect of epoch length on energy savings.

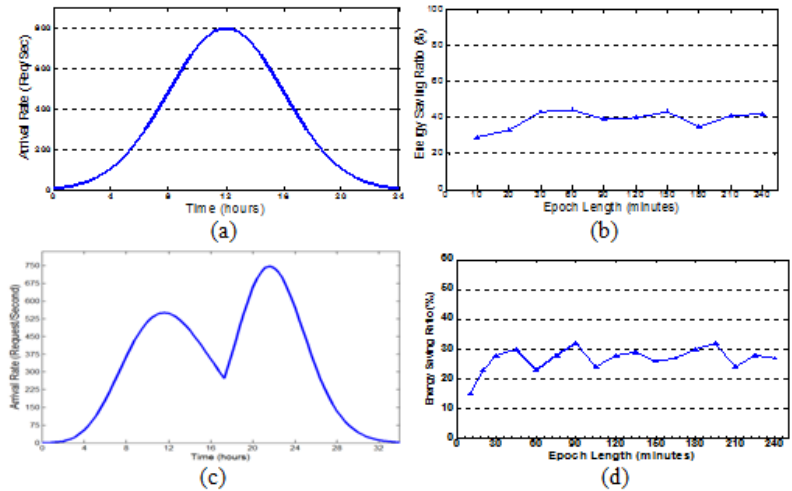

Figure 9: (a), (c): Dynamic data request arrival rate patterns; (b), (d): Effects of epoch length on energy saving ratio for (a) and (c) respectively;

Figure 9-(a) shows a dynamic arrival rate pattern in a database dynamic I/O workload used in our experiments. Figure 9-(b) shows the effect of the epoch length on the energy saving ratios for this trace. It is observed that the energy saving ratio does not fluctuate significantly under different epoch lengths more than 30 minutes. In other words, there is no observed correlation between the epoch length and the energy saving for the epoch lengths greater than the aforementioned threshold (30 minutes). As a result, the energy saving ratio is insensitive to the epoch lengths

more than 30 minutes. Figure 9-(c) shows another dynamic arrival rate change pattern related to a different dynamic I/O trace used in our experiments. Similar to the previous experiment, the epoch length is incremented by 15-30 minutes for this particular I/O workload and the corresponding power saving ratio is measured separately to monitor the effect of epoch length on energy savings. Figure 9-(d) shows that energy saving does not change significantly for the epoch length values greater than 30 minutes. In other words, similar to the previous experiment shown in Figure 9-(b), energy saving ratio is insensitive to the epoch length larger than 30 minutes. It is important to note that we captured the same trend of results on all other dynamic I/O workloads in our experiments. The observed effect is more noticeable in the traces whose arrival rate does not have remarkable changes over the time. Therefore, according to our experimental results, we confidently determine 30-minute long epoch as an efficient choice that is well-responsive to data request arrival rate changes and also exploits energy savings.

# 6. CONCLUSION

Power consumption has increased greatly in data centers. Database management system was found to be the major energy consumer in software stack of modern data centers. Among all components of DBMS, disk storage system is one of the most power-hungry constituent. Thus, we presented our research ideas on designing an energy-aware disk storage system in database servers in this paper. We improved on the limitations of the previous work. We introduced a DPM optimization model extended with MPC strategy that can be adapted to any multi-speed disk storage system. Also, a fast-solution heuristic DPM algorithm is presented that can be integrated in large-scale disk storage systems. We evaluated our proposed methods by extensive set of experimental simulations under various synthetic I/O traces based on popular TPC benchmarks.